\documentclass{Interspeech}
\usepackage{multirow}

\usepackage{times}
\usepackage{latexsym}
\usepackage{listings}
\usepackage[T1]{fontenc}

\usepackage[utf8]{inputenc}

\usepackage{microtype}

\usepackage{inconsolata}

\usepackage{graphicx}
\usepackage{xcolor}

\title{TinyClick: Single-Turn Agent for Empowering GUI Automation}
 \interspeechcameraready 

\author[affiliation={1}]{Pawel}{Pawlowski$^*$}
\author[affiliation={1}]{Krystian}{Zawistowski$^*$}
\author[affiliation={1}]{Wojciech}{Lapacz}
\author[affiliation={1}]{Adam}{Wiacek}
\author[affiliation={1}]{Marcin}{Skorupa}
\author[affiliation={1}]{Sebastien}{Postansque}
\author[affiliation={1,2}]{Jakub}{Hoscilowicz}

\affiliation{}{Samsung R\&D}{Poland}
\affiliation{}{Warsaw University of Technology}{Poland}

\email{\thanks{Correspondence to {\scriptsize\texttt{\textless k.zawistowsk@samsung.com\textgreater}} and   \\{\scriptsize\texttt{\textless jakub.hoscilowicz.dokt@pw.edu.pl\textgreater}}}}

\begin{document}
\maketitle
\def\thefootnote{*}\footnotetext{First and second author contributed equally.}\def\thefootnote{\arabic{footnote}}
\begin{abstract}
We present an UI agent for user interface (UI) interaction tasks, using Vision-Language Model Florence-2-Base. The agent's primary task is identifying the screen coordinates of the UI element corresponding to the user's command. It demonstrates very strong performance on Screenspot and OmniAct annotations, while maintaining a very small size of 0.27B parameters and minimal latency. Moreover, training needs small compute budget of 56 GPU-hours (worth about 40 USD). Relevant improvement comes from vision-specific multi-task training and MLLM-based data augmentation. We hope that decreased needs for expensive compute resources and manually annotated data will allow to facilitate more inclusive and sustainable
research of UI agents. 
\end{abstract}

\section{Introduction}
In voice assistants, we see a shift towards MLLM-based solutions. NLU and rule-based commands are replaced with agents, that can interact with the system in similar way as the user. Single-turn agents perform actions specified by the user in the UI environment, e.g. clicking on appropriate icons. Examples of such commands are "choose first item in product list", "close the application" or "open cart". The agent can perform actions based on such commands and the current GUI screenshot captured from the user's device. Currently, UI agents demonstrate moderate accuracy~\cite{rahman2024vzenefficientguiunderstanding,baechler2024screenaivisionlanguagemodelui}, despite significant computational cost.


\begin{figure}[t]
    \includegraphics[width=\columnwidth]{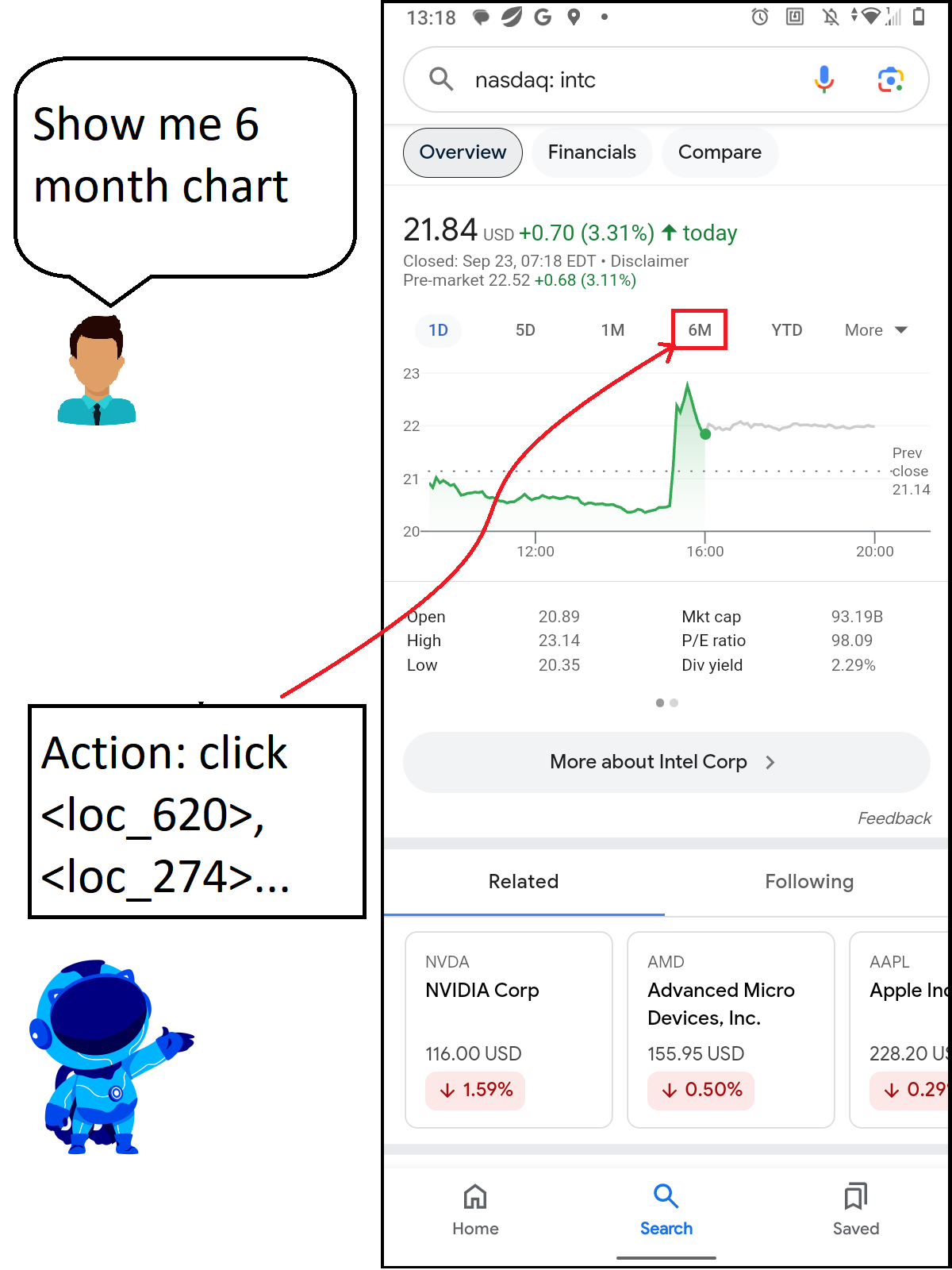}
    \caption{Example command of the downstream task. TinyClick receives a screenshot and user command and predicts bounding box of the UI element. \small Stickers from Flaticon \cite{flaticon}.}
    \label{fig:fig1}
\end{figure}

\begin{figure}[h]
    \includegraphics[width=\columnwidth]{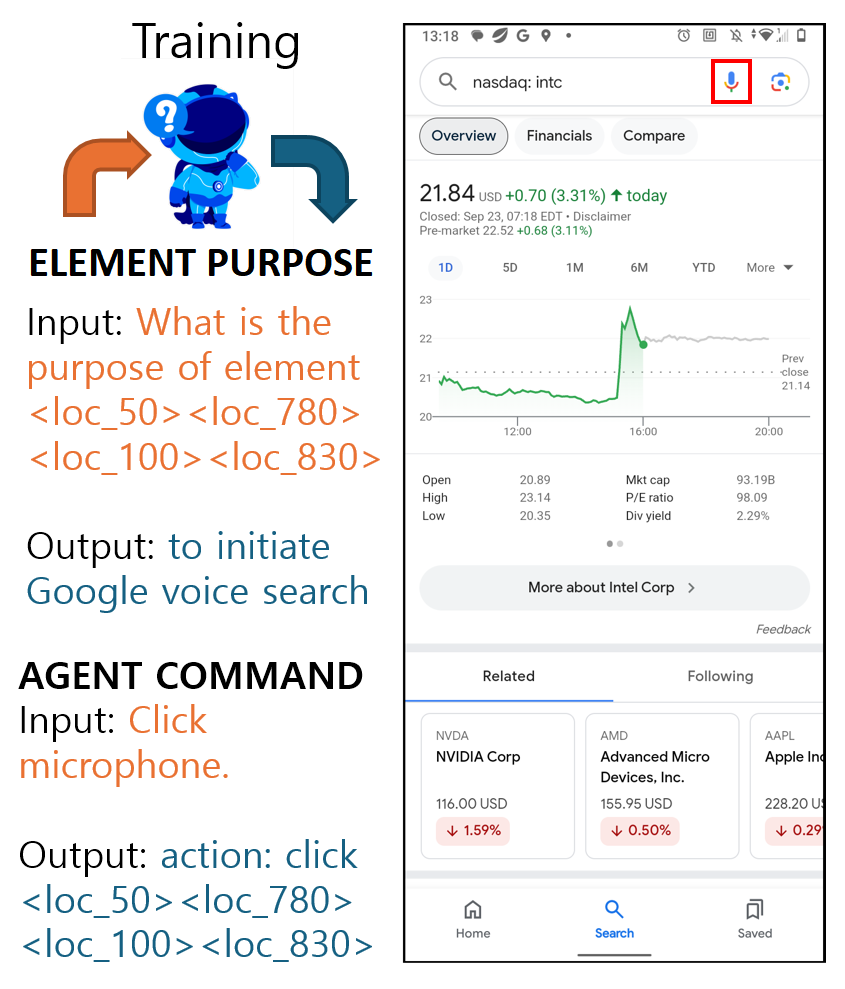}
    \caption{During the training, the model receives a question and generates an answer. Both question and answer can contain location tokens of the specific UI element. Here, the first question is about element description and the second one is a command to click specific item. Multiple different tasks can be associated with a single UI element, allowing the model to gain a better understanding of the UI. \small Stickers from Flaticon \cite{flaticon}.}
    \label{fig:fig2}
\end{figure}

We present single-turn agent based on Florence-2 model\footnote{\url{huggingface.co/microsoft/Florence-2-base}}~\cite{xiao2023florence2advancingunifiedrepresentation}, which offers state-of-the-art performance for on-device system, while having only 0.27B of parameters. When finetuning the model we investigated multiple approaches, public datasets and data preparation techniques. High-level diagrams are provided in Figure~\ref{fig:fig1} (agent inference) and Figure~\ref{fig:fig2} (agent training). Most important improvements come from multitask training. It uses multiple purpose-oriented objectives to reinforce more efficient UI representation in the model. Our key contributions are as follows:
\begin{itemize}
\item  TinyClick performs similarly to a family of much larger MLLM solutions, achieving 73.8\% accuracy on Screenspot and 58.3\% on OmniAct-Annotations.
\item  Training on vision-specific multitask data (such as generate captions and explanations) outperforms training on click-commands alone.
\item MLLM-based augmentation of corpora with commands and boxes significantly improves model performance. 
\end{itemize}
Our results suggest that extensive visual pretraining (as performed 
by Florence-2 authors) provides strong foundation for single-turn agent model, even if 
performed on the data not related to GUI and agents.

\section{Related Works}

In recent years there is much interest in Vision-Language Models (VLMs) and Multimodal Large Language Models capable of UI understanding \cite{rahman2024vzenefficientguiunderstanding, baechler2024screenaivisionlanguagemodelui, gur2018learningnavigateweb, li2023spotlightmobileuiunderstanding}.

Our focus are UI agents navigating UIs according to natural language commands of the user: such as AppAgent \cite{zhang2023appagentmultimodalagentssmartphone}, MobileAgent \cite{wang2024mobileagentautonomousmultimodalmobile}, CogAgent~\cite{hong2023cogagentvisuallanguagemodel},
Auto-UI \cite{zhang2024lookscreensmultimodalchainofaction}, V-Zen~\cite{rahman2024vzenefficientguiunderstanding}. A related problem is grounding, which involves finding UI elements corresponding to a given phrase (Ferret~\cite{you2023ferretrefergroundgranularity},
Ferret-v2 \cite{zhang2024ferretv2improvedbaselinereferring}).  Also, language-only LLM are used as agents for the Internet, using primarily HTML representation \cite{deng2023mind2webgeneralistagentweb, gur2024realworldwebagentplanninglong}. Pix2Struct \cite{lee2023pix2structscreenshotparsingpretraining} was pretrained to reproduce simplified HTML representation of screenshot (including that for masked parts of the UI). SeeClick \cite{cheng2024seeclickharnessingguigrounding}, one of best agent models so far, uses HTML data for grounding-style pretraining. Among multitask training approaches \cite{gao2024enhancingvisionlanguagepretrainingrich} utilizes 10 pre-training tasks that resemble real-world tasks. Tree-of-Lens was proposed to interpret screen content based on a user-indicated point \cite{fan2024readpointedlayoutawaregui}. 

Simultaneously with or after the initial version of this paper released in October, other models in the paradigm of SeeClick and CogAgent emerged, featuring typically large (7B or bigger) MLLM trained on UI grounding task, which produced similar or better accuracy compared to TinyClick, but with vastly larger costs. This includes \cite{xu2024aguvisunifiedpurevision,gou2024navigatingdigitalworldhumans, deitke2024molmopixmoopenweights}. Best one, AGUVIS, uses Qwen2-VL as backbone model \cite{wang2024qwen2vlenhancingvisionlanguagemodels}, which was already extensively trained on agentic use cases and rivals SeeClick in accuracy.


\section{Method}

In this work, we use Florence-2 Base~\cite{xiao2023florence2advancingunifiedrepresentation}, a 0.27B vision transformer with language modelling head trained for different vision tasks. The model latency is approximately $250$ ms, allowing for cheaper inference as well.
Vision encoder uses large 768x768 image resolution, which might be important for large screenshots. While Florence-2 uses text transformer, it is designed to handle detection and grounding tasks, encoding coordinates as single tokens. 




\subsection{Multitask Training}
Performance of transformer models can be improved by training on many related tasks. This is true of LLM~\cite{chung2022scalinginstructionfinetunedlanguagemodels}, Florence-2  \cite{xiao2023florence2advancingunifiedrepresentation}),
and models trained on UI~\cite{gao2024enhancingvisionlanguagepretrainingrich, zhang2024androidzoochainofactionthoughtgui, lee2023pix2structscreenshotparsingpretraining}.
\cite{hsieh2023distillingstepbystepoutperforminglarger, zhang2024androidzoochainofactionthoughtgui} also demonstrate training small model to predict natural language explanations of training data. Florence-2 is pretrained on object recognition, phrase grounding, captioning, segmentation and similar vision tasks. It was also trained for OCR, but not on UI data. To adapt Florence-2 for single-turn agent we investigated multitask training for UI, using tasks such as:

\begin{itemize}[noitemsep]
    \item Element captioning - generating descriptions or purposes or action expectations of UI elements based on their location on the screen.
    \item Element location - locating UI elements based on their visual description.
    \item Object detection - detection of all clickable UI elements.
    \item Agent action - locating an UI element to click or point to click based on user command.
    \item Question answering, based on screen content.
\end{itemize}
We used publicly available corpora for single-turn agent training, consisting mainly of commands and locations (bounding boxes). To prepare our training data, we used available MLLM annotations or software-based metadata and also re-annotated the data with own MLLM pipeline.
Element captioning, expectation, location and purpose were mainly based on MLLM annotations, while for object detection we used Android XML UI metadata.
This approach is not identical to \cite{hsieh2023distillingstepbystepoutperforminglarger, zhang2024androidzoochainofactionthoughtgui, xu2024aguvisunifiedpurevision, hong2023cogagentvisuallanguagemodel}, or anything that one would use with MLLM for symbolic or common sense reasoning. We do not focus on Chain-of-Though data, as the Florence2 model has no pretraining of this sort. We instead focus on augmentation of visual tasks, similarly to original Florence-2 training approach \cite{xiao2023florence2advancingunifiedrepresentation}


\begin{table}[t]
    \centering
    \renewcommand{\arraystretch}{1.2} 
    \setlength{\tabcolsep}{3pt} 
    \small
    \begin{tabular}{lc|cc}
        \toprule
         & Size & Screenspot & OmniAct-Annot. \\ \hline

        AutoUI-Base                & 0.41B   & 1.97   & 1.25                        \\
        Qwen2-VL\cite{xu2024aguvisunifiedpurevision}             & 9.6B     & 55.0$^{\dag}$      & -                        \\

                GPT-4o\cite{gou2024navigatingdigitalworldhumans}                      & -     & 18.2$^{\dag}$       & -                        \\
        SeeClick                    & 9.6B   & 55.1$^{\dag}$      & 36.8                        \\
                PaliGemma-WaveUI\cite{paligemma}                    & 3B   & 53.9      & 52.3                        \\
        \midrule
        Florence-2-base        & 0.27B   & 4.6     & -                        \\
                \textbf{TinyClick (ours)}          & 0.27B   & 73.8     & 58.3 \\ 
       \hspace{5mm}\textit{ w/o multi-task FT }              &    & 69.4    & 50.4 \\
         \hspace{5mm}\textit{ WaveUI-only}              &    & 64    & 58 \\
\bottomrule
    \end{tabular}
    \vspace{0.5pt} 
    \caption{\footnotesize TinyClick accuracy results. Goal here is choosing location to click based on user command (point prediction). The metric presented in this table is accuracy as presented in Section~\ref{sec:metric}. All values are in percentage (\%). Cited results are marked by $\dag$. Screenspot results were recalculated as mean of accuracies for mobile, desktop and web. 0.41B for AutoUI includes vision model from Blip2.}
    \label{tab:results}
    \vspace{-0.5cm}
\end{table}

\begin{table*}[h]
    \centering
     \begin{tabular}{llccccc}
      \toprule
         & & CogAgent\cite{hong2023cogagentvisuallanguagemodel} & SeeClick\cite{cheng2024seeclickharnessingguigrounding} & \emph{TinyClick} & AGUVIS-G\cite{xu2024aguvisunifiedpurevision} & UGround  \\ \hline
        \multicolumn{2}{c}{Size [B]} & 18 & 9.6 & \emph{0.27} & 9.6 & 7 \\
        \hline
        \multirow{2}{*}{Mobile} & Text & 67 & 78 & \emph{86.8} & 88.3 & 82.8 \\
        & Icon & 24 & 52 & \emph{62} & 78.2 & 60.3  \\
        \multirow{2}{*}{Desktop} & Text & 74.2 & 72.2 & \emph{90.2} & 88.1 & 82.5  \\
        & Icon & 20 & 30 & \emph{54.2} & 70.7 & 63.6  \\
        \multirow{2}{*}{Web} & Text & 70.4 & 55.7 & \emph{80.9} & 85.7 & 80.4  \\
        & Icon & 28.6 & 32.5 & \emph{56.3} & 74.8 & 70.4 \\
        \hline
        \multicolumn{2}{c}{Average} & 47.4 & 53.4 & \emph{71.7} & 81.8 & 73.3  \\
\bottomrule
     \end{tabular}
         \label{tab:results2}
             \vspace{0.5pt} 
    \caption{\footnotesize Accuracy results on ScreenSpot.
    other results than TinyClick  reported by \cite{cheng2024seeclickharnessingguigrounding, xu2024aguvisunifiedpurevision}.}
\end{table*}

\begin{table*}[h]
    \centering
    \small
     \begin{tabular}{lcccc}
        \toprule
         Dataset & Rows & Use & Type & huggingface.com link\\
        \midrule
        AMEX OD & 98.5k &   30\% & object detection (metadata) & Yuxiang007/AMEX \\
        AMEX Functionality & 296k & 50\% & commands & Yuxiang007/AMEX \\
        AMEX Purpose & 296k & 50\% & multitask (MLLM) & - \\
        AMEX Expectation & 28.9k & 100\% & multitask (MLLM) & - \\
        Wave UI & 44.2k & 100\% & commands & agentsea/wave-ui \\
        Wave UI expectation/caption/purpose & 3x44.2k & 100\% & multitask (MLLM) &  agentsea/wave-ui \\
        GUI Course web-single & 102.8k & 100\% & commands & yiye2023/GUIAct \\
        GUI Course caption/expectation & 2x51.4k & 100\% & multitask (MLLM) & - \\
        RICO ScreenQA & 56.6k & 100\% & visual question answering & rootsautomation/RICO-ScreenQA \\
        Android Control & 51.9k & 100\% & commands & GCloud, see p. \pageref{ancont} \\
        \bottomrule
     \end{tabular}
         \caption{\footnotesize Trainset composition for best training configuration (Table \ref{tab:results}).}
    \label{tab:datasets}
\end{table*}


\section{Experiments}
\subsection{Training Datasets}

We use following public datasets: WaveUI \cite{waveui} (we do not use WebUI due to license issue and remove Screenspot and OmniAct annotations, as these are benchmarks), AMEX \cite{chai2024amexandroidmultiannotationexpo}, Mind2Web \cite{deng2023mind2webgeneralistagentweb}, GUI Odyssey \cite{lu2024gui} (it is not included in our final training), GUI Course \cite{chen2024guicoursegeneralvisionlanguage}, \label{ancont} AndroidControl \footnote{Google Cloud link: \url{gresearch/android\_control}} \cite{li2024effectsdatascalecomputer}, ScreenQA \cite{hsiao2024screenqalargescalequestionanswerpairs}.
For WaveUI we use commands (for agent action task) as well as provided MLLM-generated expectation (expected outcome), purpose and captions of UI elements. We use similar approach to WaveUI to augment AMEX and GUI Course, where commands or functionalities of UI elements are provided with UI elements location. We use InternVL2-26B \cite{chen2024internvlscalingvisionfoundation} to annotate these data with purposes, captions, or expectations.
For AMEX, we use the 'functionality' field (manually annotated purpose of UI element) as well as Android XML annotations. Information about our training set that produced results in Table \ref{tab:results} can be found in Table \ref{tab:datasets}, showing that training uses 845k rows in total. 

\subsection{Benchmarks}

\label{sec:benchmarks}
We use two benchmarks: Screenspot\footnote{\url{huggingface.co/datasets/rootsautomation/ScreenSpot}} \cite{cheng2024seeclickharnessingguigrounding} and OmniAct\footnote{\url{huggingface.co/datasets/Writer/omniact}} annotations \cite{kapoor2024omniactdatasetbenchmarkenabling}. The first one contains 1200 test cases divided into 3 groups: mobile, web, and desktop. For Omniact, we do not use multi-turn commands it provides, but instead use about 3400 box-annotations with icon description provided with the dataset. 
\subsection{Results Analysis}
\label{sec:metric}
The accuracy shown in Table~\ref{tab:results} 
is calculated as the average of binary outcomes, whether predicted click point (or bounding box center) falls within the original ground-truth bounding box (1 if it does, 0 otherwise). For Screenspot, we report the arithmetic mean accuracy achieved on 3 data subgroups (see Chapter~\ref{sec:benchmarks}), according to SeeClick publication \cite{cheng2024seeclickharnessingguigrounding}.

The results show strong performance improvement over other approaches, such as SeeClick \cite{cheng2024seeclickharnessingguigrounding}, AutoUI \cite{zhang2024lookscreensmultimodalchainofaction}, and other MLLMs (see Table~\ref{tab:results}).  
The results are also stronger than publicly available PaliGemma-WaveUI-3B-896\cite{paligemma}, despite the fact, that this larger model has been trained on very similar multitask dataset, with some possible contamination with our testsets. This is also true with TinyClick trained on WaveUI alone, which beats Paligemma by 10\% on Screenspot. 
Our own, further experiments with finetuning Paligemma did not produce results, as only LoRa with small batch size was operational on 80GB A100 GPU and compute budgets needed are much larger, that those needed for TinyClick. This suggests that Florence2 pretraining on grounding and detection is important,
facilitating good results in a much smaller model and compute resources.

While Florence2 architecture is similar to that of MLLM, it differs in size and in pre-training. It has very small language modeling part, that lacks textual reasoning training (except for simple grounding phrases), but it is trained on mixture of visual task. As we follow Florence-2 approach, similar finetuning strategy adapted to GUI allows to produce our result.

\subsection{Methodology}
Large size and real world significance of the observed effect facilitates our belief in significance and credibility of the findings, especially when using much different pretrained model, than previous solutions. This follows Gosset-Neyman-Pearson methodology discussed by \cite{ioannidis} and \cite{ziliak}, which advise prioritizing strong, dominant effects. In complex systems it is easy to find many statistically significant effects, but that rarely correlates with stability and real-world significance. We, on the other hand, saw a very high gain in accuracy, with much-decreased model size and compute budget, compared to other multimodal transformers used before.
Secondly, we corroborate our claims by performing balanced experiments. Thirdly, we verified that the claims presented are supported by statistically significant results. Smallest significant ($2\sigma$) difference in accuracy is 1.8\% for OmniAct and 2.6\% for Screenspot (calculated from Beta conjugate prior). This also accounts for the observed variance of repeated experiments (about 1\% for Screenspot). Results below the significance threshold are not considered conclusive. We used 3 epochs as initial setting and tested training up to 5 epochs, which did not produce statistically significant improvement.  We trained models on 4xA100 GPU (except Table \ref{tab:results_gui_course}, which used 1xA100 and TinyClick-WaveUI that used 1xA100 with 4 epochs). Training the best checkpoint required about 56 GPU-hours (18h on 4xA100). The best checkpoint result in Table \ref{tab:results} is the average accuracy for 3 training runs. Other results were calculated for single training run. 

\subsection{Dataset insights}
We performed an ablation study, removing the numerically largest AMEX, multitask data and MLLM multitask data  (see Table~\ref{tab:results2}).
Multitask data seems more important than larger amount of commands, and MLLM data more important than metadata. AMEX produces only small gain, despite large number of examples.
Some of multitask data (as WaveUI's) can be used either as grounding objectives (generate location given phrase) and annotation objectives (generate phrase given location). The latter appears to provide stronger performance gain. To corroborate this, we used GUI Course Web single-action commands and annotated each example with expectation and caption using InternVL2-26B MLLM (method analogous to WaveUI). For training with 51k commands, 6k MLLM generated annotations of each type improved result (see Table \ref{tab:results_gui_course}), but only for annotation objective. 

Similar ablation was performed on AMEX with metadata multitask (based on Android XML annotations) compared to MLLM multitask (generated by us) and 50\% of commands (see Table~\ref{tab:amex}). All multitask approaches yield 59\% on Screenspot, but MLLM multitask is better on OmniAct, outperforming software-based metadata. We also tested MLLM-based labeling of screenshots without manual supervision; this however was much weaker than smaller high quality corpora on present benchmarks.


\subsection{Fail analysis}
30\% of failed examples suggest spurious signals that were taught to produce right answer for wrong reasons. One such example is positional bias: model clicks on the left and the top of the screen, where various menus are often found. Other is misinterpretations of similar icons. 20\% of failed examples are "missed" clicks: model clicks very closely to correct button, but slightly off-mark.


\begin{table}[h]
    \centering
    \small
     \begin{tabular}{lcc}
        \toprule
         & Screenspot & Omniact \\
        \midrule
        AMEX alone, 50\% commands          & 51.4 & 19.6 \\
        \hspace{2mm}\textit{+ metadata and MLLM multitask}     & \underline{59.6} & \underline{25.0} \\
        \hspace{2mm}\textit{+ metadata multitask}              & 59.2 & 25.0 \\
        \hspace{2mm}\textit{+ MLLM multitask}              & \textbf{59.7} & \textbf{31.2}\\
        \hspace{2mm}\textit{+ 50\% commands}              & 47.6 & 14.0 \\
        \bottomrule
     \end{tabular}
         \caption{\footnotesize Annotation and multitask training on AMEX.}
    \label{tab:amex}
    \vspace{-0.5cm}
\end{table}

\begin{table}[h]
    \centering
    \small
     \begin{tabular}{lc}
        \toprule
         & Screenspot \\
        \midrule
        TinyClick on 50\% GUI Course web single          & 52.0 \\
        \hspace{5mm}\textit{+ 6k annotations - text output}     & \textbf{55.9} \\
        \hspace{5mm}\textit{+ 6k annotations - box output}              & 53.1 \\
        \hspace{5mm}\textit{100\% GUI Course web single}              & 53.2 \\
        \bottomrule
     \end{tabular}
         \caption{\footnotesize Annotation and multitask training on GUI Course web single turn data.}
    \label{tab:results_gui_course}
    \vspace{-0.5cm}
\end{table}

\begin{table}[h!]
    \centering
    \small
     \begin{tabular}{lc}
        \toprule
         & Screenspot  \\
        \midrule
        TinyClick    & 72.5 \\
        \hspace{5mm}\textit{w/o AMEX}              & 71.9 \\
        \hspace{5mm}\textit{w/o multitask}  & \textbf{67.3} \\
        \hspace{5mm}\textit{w/o multitask except metadata}  & \underline{68.1} \\
        \bottomrule
     \end{tabular}
         \caption{\footnotesize Training data ablation study of TinyClick model. Trainset like Table \ref{tab:datasets} but only 10\% of GUI course and 10\% of AMEX OD used.}
    \label{tab:results2}
\end{table}

\section{Conclusions}
TinyClick model strongly improves present baselines, demonstrating new approach to agentic models, based on small visual model and training procedure suited to training it. It is achieving 73.8\% on Screenspot and 58.3\% on OmniAct Annotations, while being much smaller than competing solutions and small enough to work with sub-second latency. Furthermore, we confirmed that augmenting manually annotated agent data with MLLM for multitask training improves the performance.

\section{\textbf{Future Research}}
The presented contribution suggests a few new research problems to investigate. TinyClick can be used as a component for multi-turn agent (such as \cite{hoscilowicz2024clickagentenhancinguilocation}). MLLMs, thus far showing weak performance on UI control, might benefit from training strategy similar to ours.  Florence-2, as a grounding and detection model, allows to adapt out-of-domain by simultaneous use of annotated images and natural language phrases. This might produce improvements in computer vision, especially with scarce visual data. The results suggest that cheap, automated MLLM augmentation is sufficient for better performance.

With this work, we seek to answer one significant concern: provide a baseline solution that is reasonably close to commercial viability and reasonably cheap, to facilitate more sustainable, inclusive, and democratized research of GUI agents.

Lots of ongoing research in the area deals with training 7B or larger multimodal transformer on GUI automation objective. Accuracies grow with compute resources and data. However, this often has very large costs and so far produces effects that have dubious real-world utility, where not only reliability is lacking but also privacy, latency, and inference cost remain unsolved. 
In result, a proof-of-concept field of technology, such as GUI agents currently are, generates excessive financial and environmental costs, that could be cut, perhaps, by an order of magnitude.








\bibliographystyle{IEEEtran}
\bibliography{egbib}
\end{document}